\documentclass[]{aa}

\usepackage{graphicx}
\usepackage{txfonts}
\usepackage{natbib,twoopt}
\usepackage[breaklinks=true]{hyperref} 
\usepackage{array}
\usepackage{rotating}
\newcolumntype{L}[1]{>{\raggedright\let\newline\\\arraybackslash\hspace{0pt}}m{#1}}
\newcolumntype{C}[1]{>{\centering\let\newline\\\arraybackslash\hspace{0pt}}m{#1}}
\newcolumntype{R}[1]{>{\raggedleft\let\newline\\\arraybackslash\hspace{0pt}}m{#1}}

\bibpunct{(}{)}{;}{a}{}{,}             
\makeatletter
  \newcommandtwoopt{\citeads}[3][][]{\href{http://adsabs.harvard.edu/abs/#3}%
    {\def\hyper@linkstart##1##2{}%
     \let\hyper@linkend\@empty\citealp[#1][#2]{#3}}}
  \newcommandtwoopt{\citepads}[3][][]{\href{http://adsabs.harvard.edu/abs/#3}%
    {\def\hyper@linkstart##1##2{}%
     \let\hyper@linkend\@empty\citep[#1][#2]{#3}}}
  \newcommandtwoopt{\citetads}[3][][]{\href{http://adsabs.harvard.edu/abs/#3}%
    {\def\hyper@linkstart##1##2{}%
     \let\hyper@linkend\@empty\citet[#1][#2]{#3}}}
  \newcommandtwoopt{\citeyearads}[3][][]%
    {\href{http://adsabs.harvard.edu/abs/#3}
    {\def\hyper@linkstart##1##2{}%
     \let\hyper@linkend\@empty\citeyear[#1][#2]{#3}}}
\makeatother

\begin{document}

\title{Qualitative classification of extraterrestrial civilizations}

\author{
Valentin D. Ivanov\inst{1}
\and
Juan Carlos Beam\'in\inst{2}
\and
Claudio C\'aceres\inst{3,4}
\and
Dante Minniti\inst{3,5}
}

\offprints{V. Ivanov, \email{vivanov@eso.org}}

\institute{
European Southern Observatory, Karl-Schwarzschild-Str. 2, 85748 Garching bei M\"unchen, Germany
\and
N\'ucleo de Astroqu\'imica y Astrof\'isica, Instituto de Ciencias Qu\'imicas Aplicadas, Facultad de Ingenier\'ia, Universidad Aut\'onoma de Chile, Av. Pedro de Vald\'ivia 425, Santiago, Chile
\and
Departamento de Ciencias F\'isicas, Facultad de Ciencias Exactas, Universidad Andr\'es Bello, Av. Fernandez Concha 700, Las Condes, Santiago, Chile
\and
N\'ucleo Milenio Formaci\'on Planetaria -- NPF, Universidad de Valpara\'iso, Av. Gran Breta\~na 1111, Playa Ancha, Casilla 5030, Valpara\'iso, Chile
\and
Vatican Observatory, V00120 Vatican City State, Italy 0000-0002-7064-099X!ls
}

\date{Received 2 November 1002 / Accepted 7 January 3003}

\abstract 
{The interest towards searches for extraterrestrial civilizations 
(ETCs) was boosted in the recent decades by the discovery of 
thousands of exoplanets.}
{We turn to the classification of ETCs for new considerations that 
may help to design better strategies for ETCs searches.}
{This study is based on analogies with our own biological, 
historical, technological and scientific development. We take a 
basic taxonomic approach to ETCs and investigate the implications 
of the new classification on ETCs' evolution and observational 
patterns. Finally, we use as a counter-example to our qualitative 
classification the quantitative scheme of Kardashev and we consider 
its implications on the searches for ETCs.}
{We propose a classification based on the abilities of ETCs to modify 
their environment and to integrate with it: Class\,0 uses the 
environment as it is, Class\,1 modifies the environment to fit its 
needs, Class\,2 modifies itself to fit the environment and Class\,3 
ETC is fully integrated with the environment. 
Combined with the classical Kardashev's scale our scheme forms a 
2-dimensional scheme for interpreting the ETC properties.}
{The new framework makes it obvious that the available energy is not 
an unique measure of ETCs' progress, it may not even correlate with 
how well that energy is used. The possibility for progress without 
increased energy consumption implies a lower detectability, so in 
principle the existence of a Kardashev Type\,III ETC in the Milky 
Way can not be ruled out. This reasoning weakens the Fermi paradox, 
allowing for the existence of advanced, yet not energy hungry, 
low detectability ETCs. The integration of ETCs with environment 
will make it impossible to tell apart technosignatures from natural 
phenomena. Therefore, the most likely opportunity for SETI searches 
to find advanced ETCs is to look for beacons, specifically set up by 
them for young civilizations like ours (if they would want to do 
that remains a matter of speculation). The other SETI window of 
opportunity is to search for ETCs at technological level similar to 
ours. To rephrase the famous saying of Arthur Clarke, sufficiently 
advanced civilizations are indistinguishable from nature.}

\authorrunning{V. D. Ivanov}
\titlerunning{Classification of extraterrestrial civilizations}

\maketitle

\section{Introduction}\label{sec:intro}

At a fundamental level the search for extraterrestrial civilizations 
(ETCs) is motivated by scientific curiosity. We want to understand 
how intelligent life and intelligence arise and evolve, to compare
biologies, histories and social structures 
that have developed completely independently 
from each other. Undoubtedly, there are potential risk from the 
contact with an ETC \citep{2014RiskMa.16...63N}, but there are also 
indications that it may have a stimulating effect on the humanity 
\citep{2017FrontP..8.2308K}. On a long run, the transfer of new 
knowledge from fundamental sciences to industry is likely to induce 
a fast economic growth and on pure psychological level we will 
have -- for the first time -- a truly external scale to measure 
ourselves and our achievements up.

However, all these promising prospects ultimately require a successful 
search for extraterrestrial intelligence (SETI). The SETI programs in 
the last half a century have been fruitless.
One possibility is that the Universe is empty, but the commonly found 
ingredients of -- at least our form of -- life cast some doubts on 
this hypothesis.
Another option is that the ETCs are extremely rare. Therefore, the 
success is just a matter of time and increased sensitivity of the 
techniques we are already applying. 
Without listing all the possibilities for the {\it silentium universi}, 
let us consider the possibility that our search strategies may be 
wrong. Indeed, some time ago \citet{2011JBIS...64..156B} pointed at 
some caveats in our strategies: SETI mostly remains an effort isolated
from the wider astronomical and astrobiological studies, and the SETI 
proponents refuse to adopt a broader multidisciplinary approach and to
consider important criticism.

So far the dominant SETI approach, going back to 
\citet{1959Natur.184..844C}, relies (i) on the willingness of ETCs to 
be detected or (ii) on their unwillingness or failure to control their 
own energy waste (that we could detect). In the former case we are 
searching for radio beacons set up with the purpose to be visible to 
other ETCs and in the latter -- for the emission that would leak 
into space in the course of regular radio communications. The transfer 
of the searches to other ranges of the electromagnetic spectrum does 
not make a fundamental difference 
\citep{1977JBIS...30..112F,2014ApJ...792...26W}.

\citet{1964SvA.....8..217K} evaluated the feasibility of these 
approaches in radio and as a by-product developed a quantitative 
scheme to measure the stage of an ETC's advancement via the total 
amount of energy they have at their disposal. 
Undoubtedly, Kardashev's classification is still useful for SETI 
because it helps to define benchmark energy capabilities of the 
ETCs and from these to set up sensitivity requirements for the 
SETI equipment.
However, recent years brought up some new considerations. For 
example, \citet{2010AsBio..10..475B,2010AsBio..10..491B} argued 
about cost-optimized means of interstellar communication, and 
proposed some strategies to that effect \citep[see more arguments 
for cost-saving in][]{2013AcAstr.89..261D}. 
\citet{1964SvA.....8..217K} had not taken these into account. He
implicitly makes the assumption that unlimited resources are 
available to the ETCs, allowing unconstrained growth of the energy 
production and consumption, at least up to galactic scale. 
Furthermore, the estimates of Kardashev are upper limits that give
the maximum energy available for interstellar communication for 
the given level of ETC development.

Our historic and modern experience can hardly support unconstrained
growth.
Therefore, it is compelling to re-examine the ETC classification as a 
guiding tool for SETI strategies, aiming to optimize them and to arrive 
to a new priority scale for the different search methods.
The next section reviews the Kardashev's classification and its 
implications. Section\,\ref{sec:new_scale} describes a new quantitative
scale proposed here, and the final Section\,\ref{sec:summary} sumarizes
this work.

\section{Quantitative classification: Kardashev's 
scale}\label{sec:Kardashev}

\citet{1964SvA.....8..217K} introduced a classification scheme for 
ETCs based on the energy, available to them. This is a quantitative 
approach, well-justified in the context of that study, aimed to 
determine the technical feasibility of the 
communication between civilizations. He considered an isotropic radio 
emission, and estimated that transmitting with a data rate of 
3$\times$10$^5$\,bits\,sec$^{-1}$ at distance of $\sim$3\,Mpc -- this
is just below the distances to the M81 and Cen\,A groups of galaxies
and encompasses a significant number of galaxies, increasing 
significantly the number of ETCs that can potentially be detected, in 
comparison with more modest emitting power that would limit an ETC 
search to nearby stars. The estimate yielded a required transmitting 
power of 4$\times$10$^{33}$\,erg\,s$^{-1}$, comparable to the total 
solar luminosity. 

\citet{1964SvA.....8..217K} concluded that the transmitting power is 
the controlling parameter of the data rate and covered distance. This 
prompted him to build a classification of ETCs based on the energy in 
their disposal:

\begin{itemize}
\item Class\,I -- an ETC in possession of all energy of its planet
or $\sim$4$\times$10$^{19}$\,erg\,s$^{-1}$
\item Class\,II -- an ETC in possession of all energy of its star 
or $\sim$4$\times$10$^{33}$\,erg\,s$^{-1}$
\item Class\,III -- an ETC in possession of all energy of its galaxy
or $\sim$4$\times$10$^{44}$\,erg\,s$^{-1}$
\end{itemize}

The first Class is the easiest to comprehend, because it implies a
technological level close to the present-day Earth's. The humanity 
itself is approaching this level of energy consumption. Right now we 
are still limited mostly to the Earth's fossil fuel and atomic energy 
from some radioactive elements; the renewable sources of energy are 
still underutilized, but their contribution in the total energy budget 
of our civilization is increasing.

The second Class is more hypothetical. \citet{1964SvA.....8..217K} gives
as an example the Dyson sphere \citep{1960Sci...131.1667D}. Such a
structure is unstable against collapse, as pointed by many authors 
\citep[e.g.][]{1977PASAu...3..177S}. This problem can be addressed 
without calling for speculative technology or physics by breaking 
the sphere into a swarm of individual elements often called Dyson 
swarm. Each of these elements is not unlike the space habitats 
proposed by \citet{1979CosSe...1Q..16O}, but they must be quite numerous
to provide a covering factor close to unity, so nearly the entire energy
of the star is captured -- as required by the definition of the Class II
ETC. 

\citet{1960Sci...131.1667D,1964SvA.....8..217K,1966ApJ...144.1216S}
realized that the most 
prominent signature of both the ETCs' energy metabolism and of the 
Dyson sphere would be the infrared (IR) radiation and a number of 
searches for stars with IR excesses have been carried out since, mainly 
at stars on the main sequence that are long-lived and are not expected 
to show IR excess, all with negative results
\citep{1985IAUS..112..315S,1998AcAau..42..607T,2000AcAau..46..655T,2004IAUS..213..437J,2009ApJ...698.2075C,2014ApJ...792...26W,2015ApJS..217...25G,2016IJAsB..15..127O}.
Searches in the optical have also been considered by 
\citet{2018IJAsB..17..356O} who predicted anomalous variability of the 
sphere's structure due to oscillations.
\citet{2018ApJ...862...21Z} argued that a Dyson sphere with a 
covering factors less than unity can be recognized as a sub-luminous 
source, as long as an accurate parallax measurement is available. They 
searched for such object combining the {\it Gaia} Data Release 1
\citep{2016A&A...595A...1G} with the Radial Velocity Experiment 
\citep[RAVE;][]{2017AJ....153...75K} and found a few stars with lower 
intrinsic luminosity than expected for their spectral Class and with no 
detectable IR excess. However, alternative explanations such as unseen 
companions that might compromise the astrometric solutions or gray dust 
can not be fully excluded.

The last Class in the Kardashev's classification is the most speculative 
and our current technology gives little clues how an ETC could capture 
and utilize the energy of an entire galaxy. One option is a simple 
quantitative expansion of Class\,II ETCs, populating a galaxy with 
multiple Dyson spheres, whose total number is comparable to that of the 
stars in the galaxy. Similarly to the sphere-enshrouded stars, the 
galaxy will become fainter and redder and move away from the usual
scaling relations such as the Tully-Fisher relation. Following this 
argument, \citet{2015ApJ...810...23Z} set an upper limit of $\leq$0.3\,\%
to the local Kardashev's Class\,III disk galaxies.

\citet{2014ApJ...792...27W,2014ApJ...792...26W,2016ApJ...816...17W} 
and \citet{2015ApJS..217...25G}
searched for ETCs with large energy supplies, mostly by means of the
Wide-field Infrared Survey Explorer \citep[WISE;][]{2010AJ....140.1868W}.
They identified some unusual objects, none of them fully matches the 
expected signatures of Class\,III ETCs. The authors converted the 
obtained observational limits into limits on the ETCs' energy supply.
Other teams also failed to detect Class\,III ETCs 
\citep{1999JBIS...52...33A,2015A&A...581L...5G,2017IJAsB..16..176O}

A possible explanation for the lack of detections is suggested by 
\citet{2019PASP..131b4102L}, who investigated the observational 
consequences if only a fraction of the stars is enshrouded. Presumably 
it is easier to build Dyson spheres around low-mass stars than around 
hot high-mass stars for which the habitability zone is further out. 
The model predicts no detectable effects if the limit is close to the 
Solar Class stars; it must be raised up to $\sim$30\,L$_{Sun}$ to make 
the presence of Dyson spheres apparent.

\citet{2016arXiv160407844L} considers an alternative to the classical
Dyson spheres -- enshrouding the entire galaxies with artificial dust 
that would turn them effectively into black boxes, bright only in 
the microwave spectral region. He searched the {\it Planck} Catalog of 
Compact Sources \citep{2016A&A...594A..13P}, with negative result. 

Although the last two Class are purely hypothetical, the Kardashev's 
classification gained foothold in the ETC studies because of its 
convenience and the straightforward quantitative parameterisation. In 
his excellent review, \citet{2015SerAJ.191....1C} shows with multiple 
examples that the scale had a strong effect on the many SETI searches 
over the last five decades, on the strategies that these projects have 
adopted, and on the interpretation of their results.

\section{Qualitative classification of ETCs}\label{sec:new_scale}

The main motivation to re-examine the existing ETC classification is
the question how a hypothetical ETC would use the available energy
beyond the somewhat brute force approach of emitting it in space or 
blowing things up and building artificial space habitats. In practical 
terms, we propose to measure this quality of use as the level of 
interaction with the Universe. 

We can turn to the humanity's own scientific and technological 
progress, to trace the capabilities to manipulate matter: mechanical -- 
chemical -- atomic -- nuclear -- etc. One can only speculate what the 
next levels will be -- \citet{1997RvMP...69..337A} mention the
annihilation of CDM particles as a possibility. This is similar to 
the ETC classification scheme of \citet{1999ilss.book.....B}, who
uses as metrics the level of manipulation of the microworld. 
However, we can generalize further, combining these interactions 
into a single process -- of modifying our environment. The humanity 
entered this stage the moment the first tool was used. 

From this prospective the next step will be to start modifying 
ourselves, to match the environment. The modern medicine is on the 
verge of this transition -- from curing organisms to upgrading them. 
It is one step form the gene therapies that prevent a fetus from 
developing some dangerous diseases to improving it. Indeed, the 
CRISPR-Cas9 (clustered regularly interspaced short palindromic 
repeats) technique for gene editing has recently been improved to 
allow simultaneous editing of multiple genes
\citep{2012Sci...337..816J,2019Sci...365...48S,2019NatMet.16..887C}, 
bringing both medical and commercial applications of gene therapies 
within closer reach.

Is modifying ourselves an improvement over modifying the 
environment? -- Yes, for a number of reasons. First, because we, as 
a product of semi-random evolution, are far from optimal for all the 
environments, even here on the Earth. We have evolved for a short
life in small groups, in the savannah. As a result, our brains 
have insufficient computing capabilities for the modern life when 
we have to complete complex tasks that require functioning within 
large diverse groups. One of the unfortunate consequences is that 
resort to typecasting -- a major reason for the problems we face 
with various biases in the connected global village of today 
\citep[][and the references therein]{2011dktf.book..499K}. Vast 
areas of Earth's surface near the poles and the oceans are 
marginally accessible to us.

Next, our bodies wear out quickly and by age of 50-60 we start 
facing problems with such basic components as bones, vision 
and hearing. As of 2011 about 0.2-0.3\% of people need hip joint 
replacement and 0.1-0.2\% -- knee replacement at some point of 
their lives \citep{2014DAI...111..407W}, and those numbers are 
increasing \citep{2015JBJS...97.1386K}. In some countries eight 
in ten people wear glasses by age of 20 and the fraction has been 
rising for ages, correlating with reading and education 
\citep{2015Ophta.122.1489M,2018PRER...62..134M}. These are just a 
few easy problems, we are not discussing here the most serious 
ones such as cancer and various genetic disorders.

A major argument for improving ourselves is to boost our 
adaptability -- an important advantage in a world of nearly 
infinite environmental variety. We can not tolerate the entire 
range of temperatures and pressures without major protection 
measures even on our home planet, let alone live on any of the 
other planets in the solar System or potentially -- on any 
exoplanet.

Last but not least, adopting the strategy to modify ourselves 
removes the need to achieve consensus about how the environment 
could be changed and our civilization has a remarkably poor 
record on agreement, as the two world wars in the last century 
demonstrate.

We can bracket these two stages of ETC evolution. At the low-end, 
we extent of the term civilization to include wild animals 
that generally use the environment as it is. However, this is not
always the case. First, in a broad sense any animal modifies the 
environment -- e.g. just because of its metabolism, and second, 
there are well known examples of animal tool use
\citep[][among others]{1964Nat...201.1264G,1971ASB....3..195V,1974JHB....3..501M}
which underlines the point that the new classes that we are about 
to introduce are not discrete bins but represent a part of a 
continuous sequence.

At the high-end side we speculate that the boundary between the 
environment and one's self would eventually be diluted in the 
process of self-improvement to the point of merging the two. 
This is a natural consequence if we assume that the ultimate 
goal of the intelligence is to spread, which in more speculative 
terms may imply converting all the matter in the Universe into 
thinking matter \citep[but see][for a reasoning why an advanced 
civilization may prefer to stay dormant during the present 
cosmological era]{2017arXiv170503394S}.

To underline, we use here the level of interactions with matter 
and the degree of integration of ourselves with the environment 
as near synonymous, because the latter follows from the former:
historically, once our technological capabilities allowed it, we 
tried to modify the environment, e.g. moving from natural caves 
to purpose build housing; we are already willing to accept 
modifying our kind as long as it is seen as upgrading, even 
if it is ethically questionable -- the tendency for selective 
abortions of female fetuses in East Asia proves it 
\citep{2011CMAJ..183.1374H}.

Summarizing, we propose a new ETC classification scheme, containing 
the following three categories:
\begin{itemize}
\item Class\,0 : the environment is used as is (animals)
\item Class\,1 : we modify the environment (clothes, buildings)
\item Class\,2 : we modify ourselves to fit the environment 
(genetically improved humans)
\item Class\,3 : we merge with the environment, converting the 
dead matter in the Universe into thinking matter
\end{itemize}

Throughout this paper we denote the new classes with Arabic 
numerals, including fractional classes such as 0.5, 1.3, 2.8, 
etc. For clear separation for 
the Kardashev classes we use Roman numerals, although there are
strong arguments for fractional Kardashev classes as well 
\citep[see the discussion in][]{2015SerAJ.191....1C}.

The proposed new ETC scale is less strict than the classical 
Kardashev's scale, as the example given above of tool use by 
animals suggests. Furthermore, some of the modifications to the 
environment that we apply right now can also be interpreted as 
modifications of ourselves -- although this example is far from 
the genetic manipulations mentioned earlier, a hand watch and a 
pair of binoculars are modifications of the environment, but 
they can also be though of as removable implants aimed to improve 
the internal time keeping and the eyesight of average humans. 
However, the eye glasses or lenses despite being removable 
implants, aim to cure a disease, not to advance our capabilities 
and therefore are not indications of self-upgrade.

Some notions of the ideas proposed here can be found in the works 
of \citet{1998JBIS...51..175K,2013aste.book..633K} who considered 
a classification scheme adding to the energy resources the 
complexity of ETCs' transport, communication and other resources.
In another relevant work \citet{2014NandN..11....34N} considered 
the historic path of the humanity and concludes that it may have 
reached the point of creating an environmental utopia that 
removes the stimuli for optimal health, but stops short of asking 
the question what is the next step after that utopia -- a question 
that we address with our ETC classification scheme.

\begin{figure} 
\centering
\includegraphics[width=9.5cm]{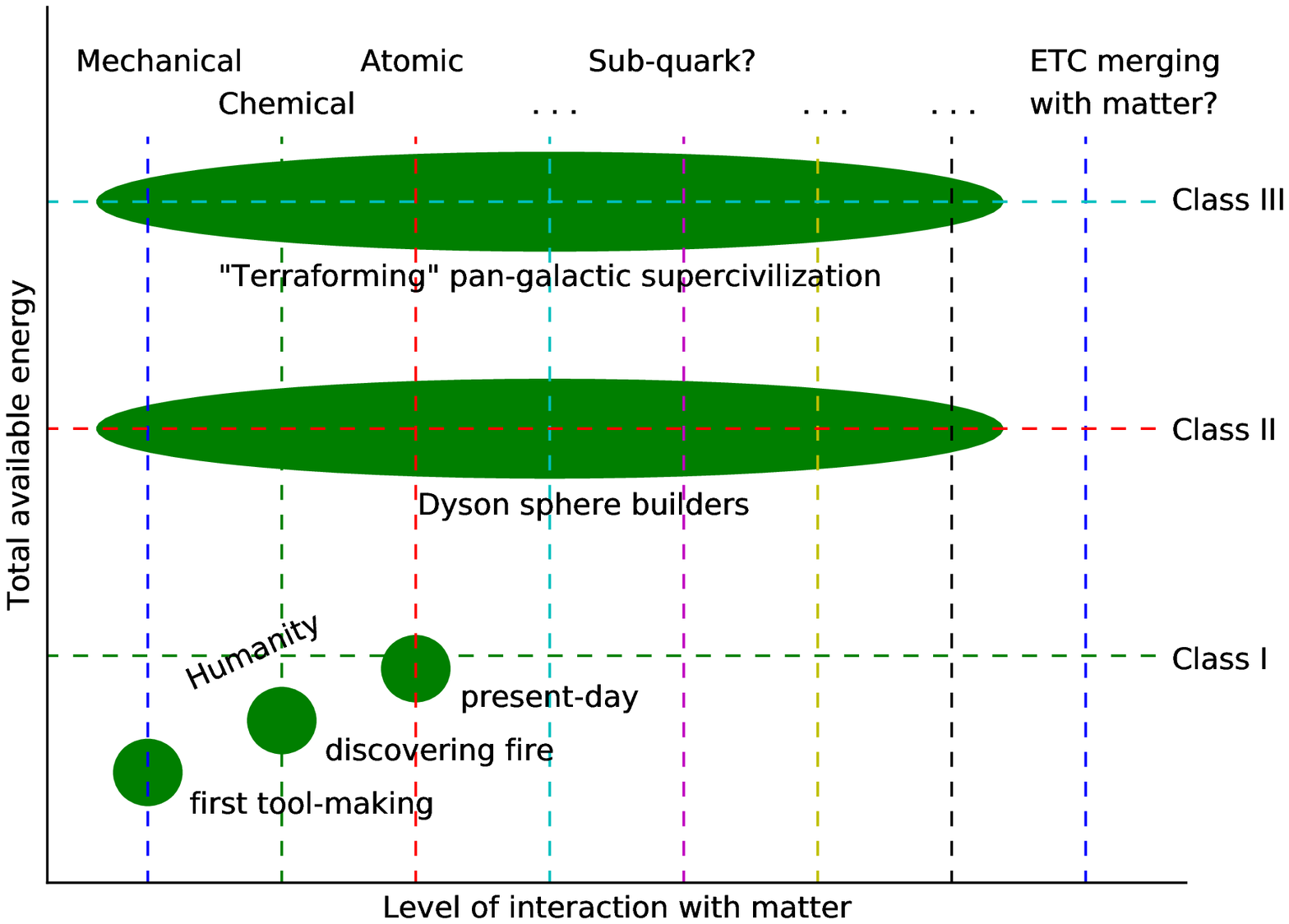}
\includegraphics[width=9.5cm]{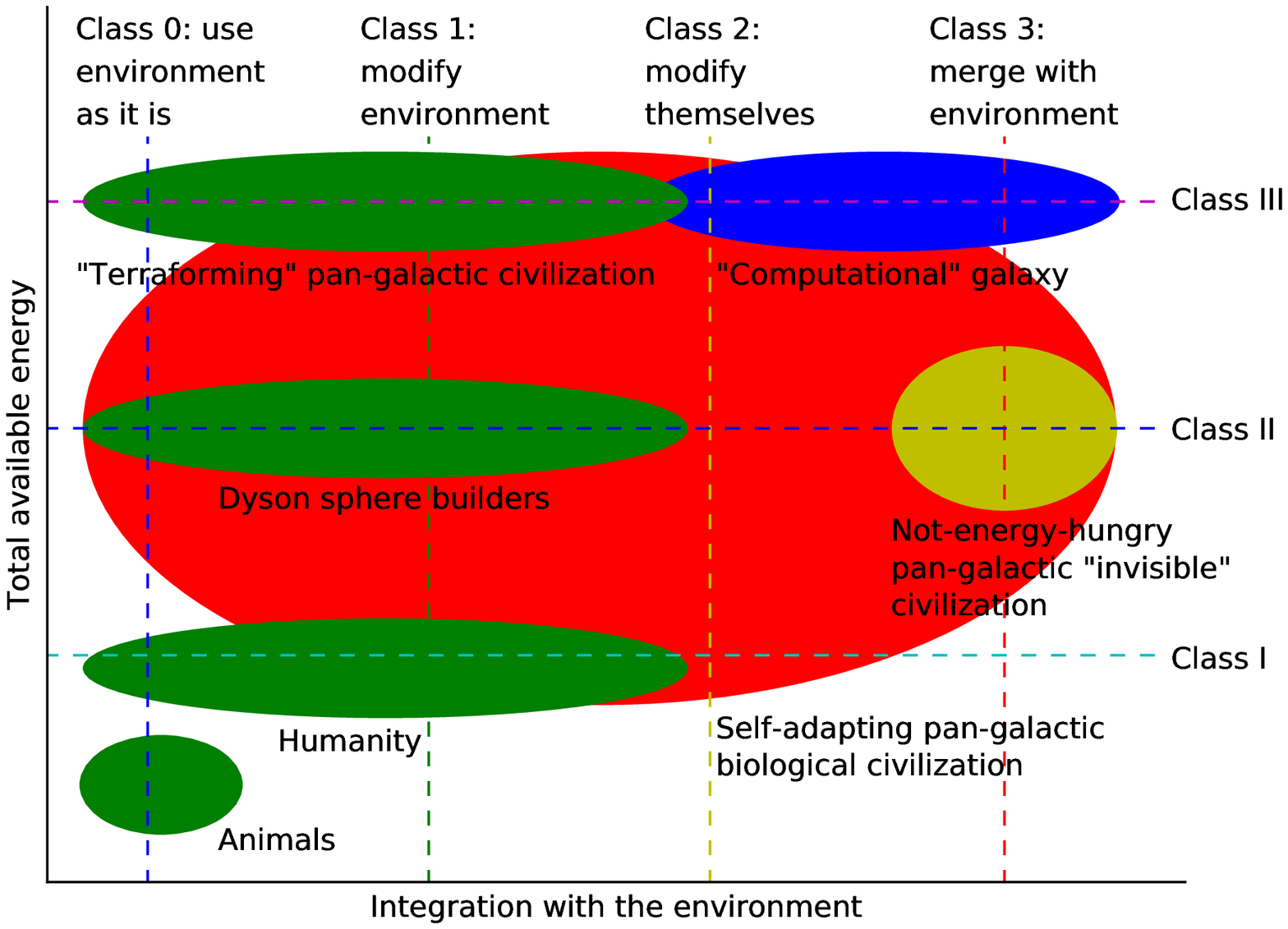}
\caption{Two-dimensional classification of the ETCs. 
{\it Top:} The horizontal axis expresses the capabilities of 
an ETC to interact with its environment.  The vertical axis 
quantifies the amount of energy available to them, as defined 
by the classical Kardashev's scale. 
{\it Bottom:} Generalized two-dimensional classification. The 
horizontal axis shows the level of integration with the 
environment.
The approximate locations of the humanity throughout history 
and of a few hypothetical civilizations are shown and labeled.
For details see Sec.\,\ref{sec:new_scale}.}\label{fig:2d_scale}
\end{figure}

The two classifications -- Kardashev's and the new one we propose -- 
can be combined to form a single 2-dimensional scheme that describes 
the ETC's progress with two parameters: the quantity of the available 
energy and the quality of its use. Figure\ref{fig:2d_scale} 
demonstrates this scheme with a few examples. The approximate 
locations of the humanity throughout history are shown on the upper 
panel: the first tool-making illustrates the mastering of the 
mechanical energy, the discovering of fire -- the widespread use of 
chemical energy, and our present day state is characterized by use 
of atomic energy and an incomplete use of the entire energy 
available on our planet. Dyson sphere building civilization and a 
conventional pan-galactic supercivilization (e.g. one that expands 
though its home-galaxy by means of multiple Dyson spheres) are also 
shown. Presumably, the two last examples have not achieved the 
level of self-modification that characterizes our Class\,3 ETC; we 
describe such civilizations with the word terraforming with 
quotation marks to underline our wider interpretation: adjusting 
the environmental condition to ones needs in general, not 
necessarily on a planetary surface. 

The bottom panel shows the Earth animals, with an offset from the 
pure Class\,0 to account for the tool usage, e.g. by chimpanzees. 
The present-day humanity spans the regions of using and modifying 
the environment, but stops short of the self-modifications. 
Barring major catastrophic events, we will probably reach that 
level of technology in the foreseeable future. Again, we mark the
loci of Dyson sphere builders and terraforming pan-galactic 
supercivilization, and we add a hypothetical civilization that has 
converted its host galaxy into a computational environments -- the
heat losses of such ETCs can potentially be detected by the WISE 
searches \citep[e.g.,][among others]{2014ApJ...792...27W}. The 
largest,
all encompassing class of civilizations is that of the adaptable, 
self-modifying ones. Changing the paradigm from physical change of 
environment to biological or even post-biological modification and
optimization of the living organisms changes the energy 
requirements. Indeed, the biological or computational research do 
not pose high energy demand. Our framework opens up a particularly 
interesting possibility -- a self-modifying civilization that does 
not need vast amounts of energy because it is fully adaptable to 
the environment.

\section{Implications for our notion of ETCs}\label{sec:implications_ETCs}

The new classification proposed here is deeply rooted on 
humanity's own evolution and may be biased in ways that can not 
be evaluated as long as we only know of one intelligent species -- 
our own. Therefore, we assume that the lessons we learn from the 
humanity's evolution and history are -- at least to some degree -- 
typical for at least some ETCs. The most obvious advantage of our 
scheme is the novel way of thinking about the ETC: we acknowledge 
that the parametric space that SETI searches need to cover can not 
be described with a single parameter as \citet{1964SvA.....8..217K} 
proposed. We introduce the question of how the available energy is 
used and what is its impact on the interaction with the matter in 
the Universe.

First, our classification scheme address in a new way the 
important question of detectability -- the ultimate strength of 
Kardashev's work that was developed exactly to address this issue 
in the particular context of radio communications. Recently, 
\citet{2019AsBio..19...28L} concluded that the probability of 
detecting advanced ETCs' technosignatures may be two orders of 
magnitude lower than of detecting biosignatures from primitive 
life. It is worth to remember that all SETI projects have 
explored an exceptionally low fraction of the Milky Way parametric 
space that can be inhabited by ETCs -- only 10$^{-21}$--10$^{-18}$ 
\citep{2018AJ....156..260W}. The realization that the 
footprint of an ETC and its detectability -- both dominated 
mainly by the energy -- may not scale up with the available 
energy, makes this estimate optimistic.

Another point, underlined by the new framework is that the two 
classifications -- Kardashev's and the one proposed here -- are 
not directly correlated. In other words, the available amount of 
energy does not necessarily mean a more sophisticated interaction 
with 
matter and closer integration with the environment. The Kardashev 
classes are separated by vast 11-12 orders of magnitude but the 
humanity -- estimated to be using still only about 70\% of the 
energy at the disposal of a Class\,I civilization 
\citep{2015SerAJ.191....1C} -- does not seem too far from reaching 
the adequate biotechnological development to improve itself and to 
integrate with the environment. This is easy to understand if we 
keep in mind that the biological research is not as energy 
intensive as the nuclear physics, as pointed above.

Furthermore, it is uncertain whether we can expect that more 
available energy would only scale up our ability to modify the 
environment. In
other words, we still lack the understanding whether the building 
of a Dyson sphere is just a matter of having more powerful mining 
equipment and heavier rockets or of some speculative technologies 
like nano-machines, self-replicators, etc \citep[see the discussion 
in][]{2013AcAau..89....1A}. In the former case the total amount of 
available energy may play a role, but in the latter -- less so, 
hereby removing any correlation between the two ETC scales 
described here.

These ideas are not entirely new among the SETI community. Indeed, 
the two-pointed arrow in his Fig.\,1 of \citet{2015SerAJ.191....1C} 
acknowledges the possibility, that the complexity is not directly 
related to the available energy.

The third direct consequence from the broader consideration of 
ETCs' properties proposed here is the invalidation of the 
obvious statement that a Class\,III ETC does not exist in the 
Milky Way \citep{1975QJRAS..16..128H}. Our searches of such 
advanced ETCs rely on the concept of detecting their heat leaks 
\citep[e.g.][]{2014ApJ...792...26W,2015ApJS..217...25G}, on the 
observational consequences from the controlled disintegration of 
galaxies for resources \citep{2003IJAsB...2..141T}, or on searches 
for megastructures \citep{2016ApJ...816...17W}. The new 
classification scheme allows for the existence of quiet advanced
civilizations that may co-exist with us, yet remain invisible to 
our radio, thermal or transit searches. The implicit underlying 
assumption of \citet{1975QJRAS..16..128H} is that the hypothetical 
ETC is interacting with the matter on a similar level as us. We can 
not even speculate if it is possible to detect a heat leak or a 
transiting structure build by an ETC capable of interacting with 
the matter at sub-quark level, but the answer is more likely 
negative and not because that ETC would function according to some 
speculative physics laws, but because such an ETC would probably 
be vastly more efficient than us controlling its energy wastes and 
minimizing its construction projects. Would such an advanced ETC 
even need megastructures and vast astroengineering projects?

It is also unlikely the on-going SETI project would successfully 
detect the Kardashev Class\,II stellivorous ETC described by 
\citet{2016AcAau.128..251V}.
Indeed, \citet{1991base.book..412H} noted that a successful SETI 
search requires a match between the technology of the transmitting 
and receiving sides.

These consideration cast some doubts on the popular pessimistic 
conclusions about the lack of ETCs 
\citep[e.g.][]{2016arXiv160407844L} (although some concerns for a 
Great Filter intrinsic to all civilizations appears to be 
still valid \citep[e.g.][]{2019IJAsB..18..445S}.

Summarizing, the new framework leads to questioning the common 
assumption that progress is equivalent to ascending the ladder 
of energy consumption from Class\,I to III, as suggested by 
\citet{1960Sci...131.1667D} even before Kardashev came up with his 
classification. Indeed, an ETC can -- as our own history shows it -- 
progress from purely mechanical modification of its environment to 
more complex manipulation on chemical, atomic, nuclear, etc. levels 
that allow it to achieve larger impacts and more importantly, 
impacts that were not possible earlier with the simpler levels 
of interaction. However, this is not necessarily accomplished by an
ever increasing energy consumption -- the biosciences show it, and 
the opposite notion is probably a bias, due to the fact that 
astronomers and physicists akin to Kardashev, Dyson and Sagan have 
been leading the SETI research, and they come armed with the idea
that progress is embodied by a more powerful accelerator or radio
transmitters.

The final and the most important consequence from the new framework 
is the weakening of the Fermi Paradox \citep{1975QJRAS..16..128H} --
if ETCs' progress does not always imply higher energy consumption 
and waste, then progress also does not imply higher detectability 
of the ETCs. This explanation of the Fermi paradox opposes the usual 
conclusion for the rarity of ETCs. In fact, they may be common, but 
the low cross-section of ours and their level of interaction with 
the Universe would account for the {\it silentium universi}.

\section{Predictions and implications for the SETI 
strategies}\label{sec:implications_SETI}

The SETI programs search for ETC's technosignatures -- traces of 
advanced technologies. On the other hand, the searches for life at 
the crossing of the modern astronomical and biological research look 
for products form the natural life cycle \citep[not necessarily 
advanced to the level of civilization, never mind how ill-defined 
this level may be;][]{2002AsBio...2..153D,2005AsBio...5..372S,2005AsBio...5..706S,2007AsBio...7...85S}.
However, the distinction between bio- and technosignatures may not 
be clear-cut. \citet{1992AcAstr.26..257R} considered hypothetical 
animals that communicate with radio waves. Indeed, the electric 
squids and rays \citep[e.g.][]{1992AcAstr.26..257R} use electricity, 
and direct electricity generation by biological systems have been 
demonstrated \cite{2016SciRep..6.25899T}. Therefore, signatures we
commonly consider part of the technological realm, may actually 
evolve naturally. For simplification we will exclude this possibility 
from the following discussion, but we remind the reader that the most 
fundamental assumptions of SETI are not simple and straightforward.

What does this new ETC classification scheme mean for the definition 
of future SETI projects? -- Wittingly or not, the searches so far 
have been fine-tuned to detect civilizations of Kardashev Types\,II 
and III, but only ones that follow the same more-is-better philosophy 
as we do. The mechanistic transfer of this power hungry reasoning 
across a range of available energy levels wider than 25 orders of 
magnitude could be why these searches fail. The searches for Dyson 
spheres and swarms, although relatively easy with the present 
technology, also seem less than promising. Somewhat more productive 
strategy may be to search for biosignatures, because that would 
accompany life regardless from the technological development
\citep{2019AsBio..19...28L}. However, as discussed earlier, we can 
not fully rely even on the biosignatures, because an advanced ETC can 
(presumably) easily modify itself to survive beyond the limitations 
of its original habitability limitations.

The possibilities of self-modification and for further integration of 
ETCs with their environment destroys the very idea of separating 
natural and artificial phenomena and by definition makes it impossible 
for us to detect with confidence any technosignatures, because -- 
rephrasing A. C. Clarke -- any advanced ETC will be indistinguishable 
from nature. This idea of indistinguishability has been discussed 
earlier in \citet{2018grsi.book.....C}\footnote{The same idea can also 
be found in the 1971 essay {\it The New Cosmogony} by Polish writer 
and philosopher Stanislaw Lem. It can be found in the collection {\it 
A Perfect Vacuum} (trans. by M. Kandel), Northwestern U. Press, 1999, 
pp. 197–227.}. This process of technological development and 
optimization is different from the natural evolution of techno-like 
signatures proposed by \citet{1992AcAstr.26..257R}.

Undoubtedly, these arguments would be well understood by ETCs that 
have attained the levels of progress discussed here. Therefore, they 
would set up beacons to emit clearly artificial signals -- on the 
condition (addressing this major question is beyond the scope of this 
work) that they still wish to communicate with less advanced 
counterparts such as us.

These arguments do not close the possibility to find ETCs at similar 
level to ours, but given the limited energy resources available to 
those ETC, such SETI programs are limited to smaller space volume 
that can be searched with any hopes for success.

Concluding, the new framework implies two strategies for SETI:
\begin{itemize}
\item Search for ETCs similar to us, for their radio radars and 
communications, for laser beacons and laser-powered interstellar 
probes, etc.
\item Search for highly advanced ETCs (that have retained interest 
in their younger/simpler counterparts) for their energy efficient 
Benford beacons, rare/unstable element/isotope doped stars and 
white dwarfs, modulated/coordinated variables, etc.
\end{itemize}

\section{Summary and conclusions}\label{sec:summary}

A classification for ETCs based on their level of interaction and 
integration with the environment is proposed. It can be combined 
with the classical Kardashev's scale to form a 2-dimensional 
scheme for interpreting the ETC properties. The new framework 
makes it obvious that the available energy is not an unique 
measure of ETCs' progress, it may not even correlate with the 
quality of use of that energy. Furthermore, the possibility 
for progress without increased energy consumption implies a lower
detectability, so in principle the existence of a Kardashev 
Type\,III ETC in the Milky Way can not be excluded. This reasoning 
weakens the Fermi paradox, allowing for the existence of advanced, 
but energy quiet ETCs.

The integration of ETCs with environment will make it impossible 
to tell apart the technosignatures from natural phenomena. Therefore, 
the only hope for future SETI searches to find advanced ETCs is to 
look for beacons, intentionally set up by them, to be found by the
backward civilizations like ours. It remains a matter of 
speculation if advanced ETCs would be interested to communicate 
with us. The other SETI window of opportunity is to search for ETCs 
at approximately our technological level.

This new proposal is not a criticism of the 
\citet{1964SvA.....8..217K}. He carried out this work with the 
specific goal to estimate the feasibility of interstellar radio 
communications and, naturally, it was used to evaluate the 
detectability of ETCs in radio. Undoubtedly, the Kardashev's scale
will continue to be important for defining the sensitivities of 
SETI searches that utilize the strategies relying on communication 
leaks or communication beacons.

\begin{acknowledgements}
We thank the referees for the comments that helped to improve the 
paper and for pointing at the number of relevant works in the 
field of biology. We thank M.S. for the helpful discussions.
D.M. acknowledges support from the BASAL Center for Astrophysics 
and Associated Technologies (CATA) through grant AFB 170002, and 
Proyecto FONDECYT No. 1170121. CC acknowledges support from 
DGI-UNAB project DI-11-19/R.
\end{acknowledgements}

\bibliographystyle{aa}
\bibliography{class_13}

\end{document}